\documentclass[12pt]{article}
\usepackage{epsfig,cmb}

\begin{document}

\heading{RESULTS FROM CAT AND PROSPECTS FOR THE VSA}

\author{Michael E. Jones} {Mullard Radio Astronomy Observatory, Cavendish
Laboratory, Madingley Road, Cambridge CB3 0HE,
UK.} {$\;$}

\begin{abstract}{\baselineskip 0.4cm 
We have produced an image of the microwave sky with $30'$ resolution in a
$2^{\circ}$ field using the Cosmic Anisotropy Telescope (CAT). Analysis of
data taken at three frequencies near 15~GHz indicates that most of the signal
is due to the CMB, with an equivalent broad-band power of $\Delta T/T =
1.9^{+0.5}_{-0.5} \times 10^{-5}$ at angular scales corresponding to
multipoles $l = 320$--500, and $\Delta T/T = 1.8^{+0.7}_{-0.5} \times 10^{-5}$
at $l = 500$--680. We are now building a more advanced instrument, the Very
Small Array (VSA), which will cover the range $l = 130$--1800 with a
sensitivity per resolution element of $\Delta T/T \sim 10^{-6}.$}

\end{abstract}

\section{Introduction}

Measurements of the anisotropy in the cosmic microwave background radiation
(CMB) in the vicinity of the expected `Doppler peaks' ($10'$--$2^{\circ}$) are
crucial for the determination of cosmological parameters such as $\Omega$ and
$H_0$, and for discriminating between competing theories of structure
formation. Most results on these scales so far have come from balloon-borne
switched-beam bolometer systems (eg \cite{cheng,clapp}) or ground-based
switched-beam heterodyne systems at dry sites (eg \cite{ruhl,netterfield}). We
have been developing centimetre-wave {\em interferometers} which offer several
advantages over these techniques. These include rejection of groundspill and
other signals that do not move with the sky, freedom from $1/f$ noise problems
in the receivers, and the rejection of atmospheric emission fluctuations,
giving the ability to operate from less extreme
sites\cite{saunders,church}. The Cosmic Anisotropy Telescope (CAT) is a
prototype interferometer designed to test the concepts and technology for a
more comprehensive imaging interferometer, the Very Small Array (VSA). In this
paper we briefly discuss interferometric imaging of the CMB, review the CAT
results and then describe the VSA.

\section{Interferometric imaging and power-spectrum estimation}

\begin{figure}
\begin{centering}
\epsfig{figure=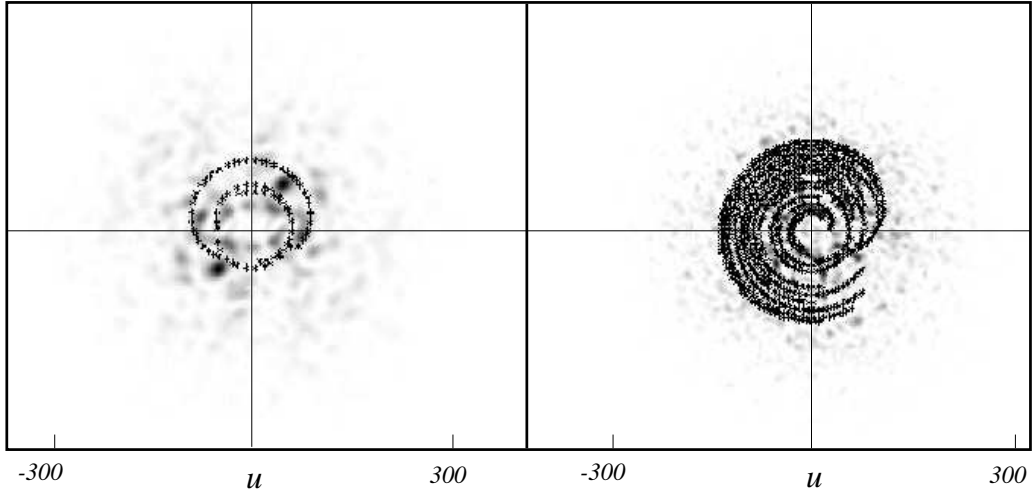,angle=270,width=15cm}
\caption{Sample 2-d power spectrum for a standard CDM model, with
aperture-plane sampling function of the CAT (left) and the VSA with
$4^{\circ}$ horns (right). Note the smaller convolution scale of the power
spectrum in the case of the VSA due to the larger primary beam.}
\label{aperture}
\end{centering}
\end{figure}

Each antenna pair of an interferometer measures one Fourier component of the
part of the sky seen by the envelope (or primary) beam of the antennas,
i.e. if a sky brightness distribution $T({\bf s})$ is observed with an
interferometer of baseline ${\bf u}$ (in wavelengths) then the measured
quantity is the {\em visibility}
\begin{equation}
V({\bf u}) = \frac{2k}{\lambda^2} \int T({\bf s}) B({\bf s}) {\rm e}^{i {\bf s.u}} \; {\rm d}{\bf s},
\end{equation}
where $B({\bf s})$ is the power reception pattern (primary beam) of the
antennas, and the factor $2k/\lambda^2$ converts from temperature to flux
density (the units of an interferometric map being Jy~beam$^{-1}$). Thus a CMB
interferometer measures (almost) what the theorist wants to know---$|V({\bf
u})|^2$ is proportional to the two-dimensional power spectrum of the sky, but
convolved with the square of the Fourier transform of the primary beam. This
places a limitation on the resolution with which the power spectrum can be
obtained. With a fixed number of antennas of a given aperture, there is a
maximum area of aperture plane that can be covered; this fixes the range over
which the power spectrum can be measured, or equivalently, the number of
resolution elements in the image. A sparsely sampled aperture plane (e.g. the
CAT, see Fig. \ref{aperture}) can be inverted to yield a map, but the poor
sampling in the aperture leads to long-range correlations in the image. This
can be ameliorated by deconvolving the image, using for example the CLEAN
algorithm, but such an image must be treated cautiously, since this is
equivalent to interpolating the visibilities into regions of the aperture
where none were measured. However, a well-sampled aperture plane can be
inverted to yield a faithful image of the sky, subject to the resolution limit
imposed by the maximum baseline, and the low spatial-frequency cut-off of the
shortest baseline.

\begin{figure}[t!]
\begin{centering}
\epsfig{figure=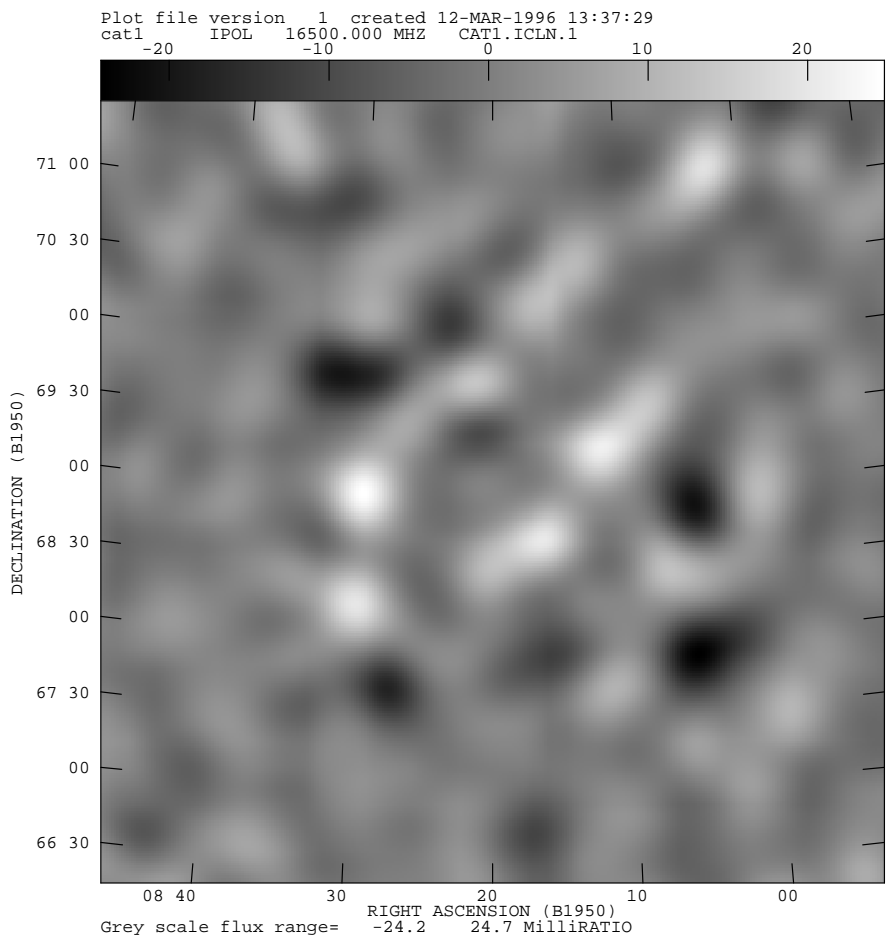,width=12cm}
\caption{Combined map of 13.5-, 15.5- and 16.5-GHz CAT data, weighted as $\nu^2$
to maximise the CMB component. The map has been CLEANed to remove the
long-range correlations due to the sparse sampling of the aperture plane. The
greyscale units are mJy~beam$^{-1}$}
\label{cat-cmb}
\end{centering}
\end{figure}

A image contains both amplitude and phase information; however, if the sky is
Gaussian the phases of the Fourier components are random and all the
information is contained in the one-dimensional power spectrum. This can
na\"{\i}vely be obtained by averaging the visibilities, first radially into
independent bins, then azimuthally, squaring them, and subtracting the
variance of the noise from each bin. However, in the case of low
signal-to-noise this can result in negative estimates for the power spectrum
in some bins. A better method is to calculate the covariance matrix expected
of the data under different power-spectrum models, and adopt a
maximum-likelihood approach to finding the best model\cite{hobson,scott}.

\section{CAT}

The CAT\cite{robson} is a three-element interferometer operating at 15~GHz,
with a primary beam of $2^{\circ}$ FWHM and a resolution of $\sim 30'$. The
three conical horn-reflector antennas provide a primary beam with sidelobes
less than $-60$~dB, minimising pick-up from the surroundings and from bright
astronomical sources. The antennas are steered in elevation by rotating the
reflectors, allowing the cryostats containing the HEMT receivers to remain
fixed relative to the horn and the turntable, which provides the azimuth
drive. The whole telescope is contained within a 5-m-high earth bank lined
with aluminium, so that all stray ray paths are reflected on to the sky. The
signals are down-converted at the turntable and sent via coaxial cable to a
control hut $\sim 100$ m away, where the 500-MHz-bandwidth IFs are correlated
using an analogue phase-switched correlator. Both linear polarizations are
used, but the design of the antennas means that the polarization of each
channel rotates on the sky as the telescope tracks.

\subsection{Source subtraction}

The most important contaminating signal for the CAT is discrete extragalactic
radio sources. We select fields at 5~GHz (the highest available frequency with
near all-sky coverage) to have minimum source content, but sources still
contribute many times the total flux expected from the CMB aniotropies. It is
therefore essential to observe these sources at higher flux sensitivity and
higher resolution, and at close to the same frequency. This is done using the
Ryle Telescope (RT). The RT compact array has five 13-m diameter antennas,
giving a flux sensitivity at 15.4~GHz of $200 \; \mu$Jy in 12~h over a $6'$
FWHM field of view, with $30''$ resolution. The flux sensitivity at each of
the three CAT frequencies described in the next section is 7~mJy; the RT can
map $(0.5^{\circ})^2$ with a noise level of 1~mJy in 12 h. Therefore in $16
\times 12$ h the RT can map the entire CAT field with sufficient sensitivity
to remove all sources to well below the CAT sensitivity.

\subsection{CAT results}

First results at 13.5~GHz\cite{osullivan} for a field centered at $08^h 20^m
+69^{\circ}$ (B1950) (the `CAT1' field), showed evidence for a fluctuation
level, after subtraction of discrete sources, of 18~mJy beam$^{-1}$,
equivalent to temperature fluctuations of about 35 $\mu$K rms. Subsequent
observations at 15.5 and 16.5~GHz\cite{scott} allow us to estimate the
foreground Galactic component and the level of the CMB signal.

The aperture-plane coverage of the CAT results in two independent radial bins,
centred at multipole numbers $l=410$ and $l=590$. We use a maximum likelihood
technique, modelling the CMB and Galactic signals as independent Gaussian
signals with variable powers in the two bins, with a fixed flux spectral index
of 2 for the CMB and variable in the range [$0,-1$] for the Galaxy. We then
calculate the expected correlation matrix of the visibilities for each model,
ranging over the five-dimensional parameter space, and find the maximum
likelihood with respect to the data. Marginalising over the Galactic
parameters gives the best estimate of the CMB power in the two bins. The data
are consistent with most of the signal at 16.5~GHz being due to the microwave
background. Taking the square root of the power, the results are $\Delta T/T =
1.9^{+0.5}_{-0.5} \times 10^{-5}$ for the bin covering the range $l = 410 \pm
90$, and $\Delta T/T = 1.8^{+0.7}_{-0.5} \times 10^{-5}$ for the bin at $l =
590 \pm 90$. Given these results, we can combine the three maps at different
frequencies, weighted as $\nu^2$, to produce an image that is mostly CMB
fluctuations (Fig. \ref{cat-cmb}). This image, unlike the one in \cite{scott},
has been CLEANed to remove some of the long-range correlations and allow the
eye to see more easily the regions of highest and lowest temperature (but note
the caveats in section 2 above).

In conjunction with other CMB results, these points provide evidence for the
existence of the first Doppler peak, and hence place constraints on the
cosmological parameters $\Omega$ and $H_0$ (see Hancock, and Rocha, this
volume).

\begin{figure}[t!]
\epsfig{figure=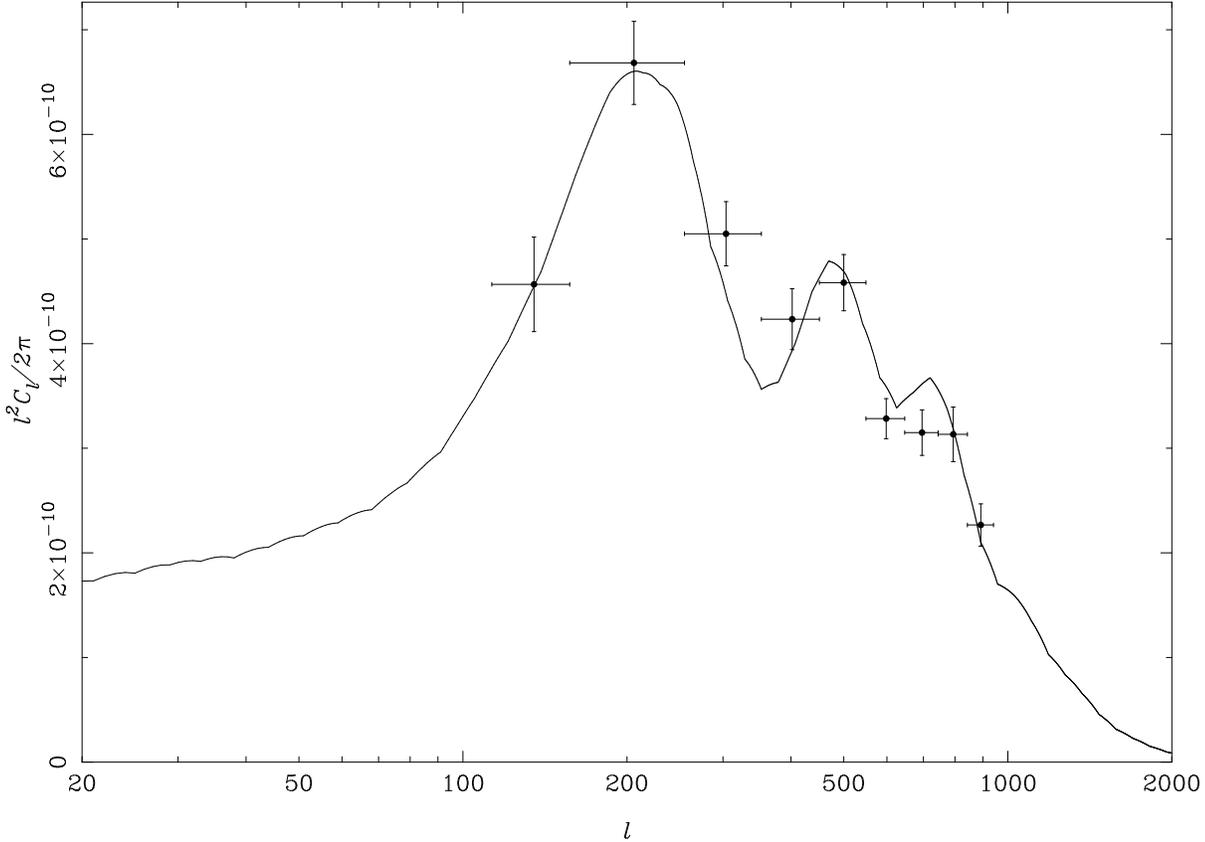,width=12cm,angle=270}
\caption{Simulation showing the ability of the VSA to recover the Doppler
peaks of the CMB power spectrum. The solid line is a standard CDM power
spectrum with $H_0 = 50$, $\Omega_{\rm b} = 0.03$ and $n = 1$, normalised to
COBE. The points are the recovered power spectrum from simulated observations
of twelve $4^{\circ}$ FWHM fields, representing one year's VSA
observations. The simulation includes the effects of removal of discrete
sources and diffuse galactic foregrounds. The horizontal error bars represent
the width of the independent bins sampled in $l$-space.}
\label{ps}
\end{figure}

\begin{table}[b!]
\small
\begin{tabular}{|l|c|c|c|c|c|c|c|}
\hline
 &          &Number of &          &            & Temperature  & Flux & Statistical \\
 & Frequency& antennas $\times$  & Field-of & Reso- & sensitivity  & sensitivity& sensitivity \\
 &  range (GHz) & polarizations   & -view    &   lution   & in 300 h     &
in 300 h &  \\
\hline
CAT & 13.5--16.5 & $3\times 2 $ & $2^{\circ}$ & $30'$ & $35\;\mu$K & 7 mJy& $9\;\mu$K \\
VSA & 26--36     & $15\times 1$ &$4(2)^{\circ}$ & $30(15)'$
& $7\;\mu$K & 5(1.3) mJy &$0.7\;\mu$K \\
\hline
\end{tabular}
\caption{Comparison of specifications of the CAT and the VSA. The
temperature sensitivity is per pixel after 300 h observation (as in
each of the three CAT maps in [8]). The statistical sensitivity is the
temperature sensitivity divided by the square root of the number of
independent pixels. Numbers in parentheses for the VSA are for the
second array with larger horns.}
\end{table}

\section{The Very Small Array}

The CAT is the prototype for the Very Small Array, which is designed to make
high-quality images of the CMB. The VSA, which has just received funding
(March 1996), will have 15 antennas and a 2-GHz bandwidth analogue correlator,
using exactly the same technology as has been successfully tested in the CAT;
the specifications are given in Table 1. The operating frequency will be in
the range 26--36~GHz, decreasing the effect of discrete sources and Galactic
emission compared with the CAT frequency range, but increasing the level of
atmospheric fluctuation emission. To alleviate the latter effect, the VSA will
be sited at the Teide Observatory in Tenerife, alongside the Jodrell
Bank--Tenerife switched-beam experiment. Observations with the Tenerife
experiment, and a prototype 33-GHz interferometer (R. Watson, priv. comm.)
indicate that the atmospheric conditions at Teide will allow the VSA to
operate essentially unhindered by the atmosphere for $> 50\%$ of the time.

Source subtraction will be as important for the VSA as for CAT; again, the
Ryle Telescope with its higher resolution and flux sensitivity will be
vital. Since the RT observing frequency is a factor of 2 lower than the VSA,
the RT will have to survey the VSA fields to a flux sensitivity a factor of
$2^{\alpha}$ better than the flux sensitivity of the VSA, where $\alpha$ is
the maximum expected spectral index of sources in the field. Once all the
sources which might be significant at the VSA frequency have been found at
15~GHz, measuring their fluxes at $\sim 30$~GHz will require only a very short
time on a large telescope, e.g. the Bonn 100-m.

The VSA will operate consecutively with two sets of horns; one set of $\sim
15$~cm aperture giving a $4^{\circ}$ beam, and a second set of $\sim 30$~cm
aperture giving a $2^{\circ}$ beam. Using the larger horns will allow the VSA
to extend to higher angular resolutions without compromising the filling
factor of the aperture plane. The CAT, with only three baselines, samples the
aperture plane very sparsely, giving only two independent points in the
one-dimensional power spectrum. The VSA will measure $\sim 100$ independent
points in the aperture plane, giving $\sim 10$ independent points on the 1-D
power spectrum (see Fig. \ref{ps}). The width in $l$ of each bin is fixed by
the primary beam, i.e. the antenna diameter; the range of $l$ is fixed by the
range of baselines possible with 15 antennas and the requirement to sample the
aperture plane uniformly. The resolution and range of the measured power
spectrum can be changed by changing the size of the antennas. With these two
arrays we will be able to measure the power spectrum from $l = 130$--1800 with
a resolution $\Delta l = 100$ at low $l$ and $\Delta l = 200$ at high $l$.

\acknowledgements{Thanks to Mike Hobson and Klaus Maisinger for the VSA
power-spectrum simulations. CAT and the VSA are funded by PPARC.}

\vfill
\end{document}